\documentstyle[preprint,aps]{revtex}
\tightenlines
\begin{document}
%\draft
\title{Local Polariton States in Polar Crystals with Impurities}
\author{Lev I. Deych and Alexander A. Lisyansky}
\address{ Department of Physics,
Queens College of City University of New York, Flushing, NY 11367}

\date{\today}
\maketitle

\begin{abstract}
We show that an impurity embedded in an ionic crystal can give rise to a novel kind of
local states. These states exist within a polariton gap of a material and are a mix of
excitations of the crystal, such as phonons or excitons, and the transverse electromagnetic field.
Electromagnetic component of the states along with the corresponding excitations of the material
are localized in the vicinity of an impurity.
\end{abstract}
\narrowtext

Impurities or other defects in otherwise  ideal crystal lattices have an considerable affect  on dynamical
properties of host crystals.  One of the most remarkable dynamical effects caused by defects
is the occurrence of local states within frequency regions forbidden for excitations of the host
materials. These states are well known for lattice vibrations
\cite{phonon1,phonon2,phonon3,phonon4}, excitons \cite{exciton1,exciton2,exciton3},
magnons \cite{magnon1,magnon2} and other kinds of excitations in periodic crystals.  Such 
local or bounded states had not been found, however, for  electromagnetic waves in ideal
crystals with isolated impurities. The reasons for this are quite clear. For local states to appear
there must be either bandgaps in the spectrum of  excitations of the ideal structure or an upper limit
of frequencies available for them. Besides that, the excitations and defects must be able to interact effectively in order to form a bounded state.   
Wavelength of light in the range of frequencies up to ultraviolet is much greater than the crystal lattice constant and characteristics scales of any possible microscopic defects. Therefore, neither of these conditions are fulfilled for
electromagnetic waves at optical frequencies propagating  through regular crystals. Bounded photon
states were eventually created in so called photonic crystals \cite{localreview}. These are
periodic
arrays of macroscopic pieces of materials with a lattice constant
of the order of a wavelength of light.  A local distortion of
the periodicity in such structures can give rise to local photon states with frequencies within a
gap between different photonic bands \cite{Yablonovitch,Joannopoulos}.

Inspite all the obvious arguments regarding infeasibility  to create local  states of
transverse electromagnetic field in convential crystals with microscopic defects, we show in this Letter that it is, nevertheless, possible. 
Our idea is to consider an interaction of an impurity with electromagnetic waves in a region of
a polariton resonance, where
light propagates as a mixed states with other excitations. Scattering of {\it extended}
polaritons due to impurities was studied in Refs. \cite{Maradudin,Mills,polarlocal}.  We  consider {\it local polariton states} , which appear, due to defects,
{\it inside a forbidden gap} between different polariton branches (reststrahlen region).  We show that dipole active isolated
defects  can form coupled states with polaritons. If a frequency of this complex falls  within the gap both electromagnetic and material components of these states are confined around the
impurity, so that they can be called local polaritons. 

In this Letter we consider a simple model of a scalar harmonic  excitation of a crystal
(polarization wave) interacting with the electromagnetic field, which is also treated in the scalar
approximation. A single defect embedded in the structure is assumed to affect the diagonal
part of  the corresponding Hamiltonian only, which means that the defect and host atoms differ
only
in their individual characteristics (e.g., masses for phonons or single atom excitation energies for
excitons). Atom polarizability of the defect is assumed to be the same as for host atoms. This assumption makes our model of the defect considerably different from that considered in Ref.\cite{Mills}, where defects were assumed to differ from host atoms only in their polarizabilities. We will show that effects due to elcetrical defectiveness are negligible comparing to those resulting from difference in mechanical (or electronic) characteristics. 

The  general results presented here do not depend upon the specific nature of the
crystal excitations (which we will call polarization waves) and when discussing them we will refer
to both excitons and to phonons.
We believe that this model reflects the most important
characteristics of the phenomenon and will serve as a good starting point for more realsitic
consideration. 

The electric component, $E$, of the transverse electromagnetic field interacting with scalar
classical polarization waves obeys Maxwell's equation
\begin{equation}
\label{Maxwell}
-\frac{1}{c^2}\frac{\partial^2 E({\bf r})}{\partial t^2} + \nabla^2 E({\bf r}) =
\frac{4\pi}{c^2} \frac{\partial^2 P({\bf r})}{\partial t^2},
\end{equation}
where $c$ is the velocity of an electromagnetic wave in the medium. The term on
the right hand side of Eq.\  (\ref{Maxwell}) is responsible for the interaction with
lattice excitations due to their polarization, $P$.
The equation of motion for the polarization waves can be written as
\begin{equation}
\label{Dynamics}
\frac{\partial^2 P({\bf r})}{\partial t^2} + \sum_{{\bf r}_1}^{}
L({\bf r}-{\bf r}_1)P({\bf r}_1) - \gamma\frac{\partial^2 P({\bf r})}{\partial t^2}
 \delta_{{\bf r r}_0} =
\frac{d^2}{4\pi}E({\bf r}).
\end{equation}
This equation describes one branch of excitations of
a crystal with a single impurity at a site ${\bf r}_0$ interacting
with the electric field $E$. The last term at the left side  
of Eq.\  (\ref{Dynamics}) represents a local perturbation due to
the impurity, where the parameter $\gamma$ accounts for a difference between an impurity and
host atoms. For phonons, $\gamma$ is the relative difference in masses, for excitons, 
it is the relative difference in atom's excitation energies.  The
dynamical matrix of a crystal, $L({\bf r}-{\bf r}_1)$, which describes an interaction between
atoms at different sites, is assumed to
be unaffected by the impurity, parameter $d$ is an effective coupling parameter of the
excitations with the electromagnetic field.  

Polaritons arise as the simultaneous
solutions of Eqs.\  (\ref{Maxwell}) and (\ref{Dynamics}). In order to derive an equation
describing the electromagnetic component of polaritons, one needs to eliminate the
polarization, $P(\bf r)$, from Eqs.\ (\ref{Maxwell}) and (\ref{Dynamics}). This can be done by
using the Green's function of Eq.\  (\ref{Dynamics}), which has the form
\cite{phonon2,phonon3,phonon4}
\begin{equation}
\label{Green}
G({\bf  r},{\bf r'}) = G^0({\bf  r} - {\bf r}') -
\gamma\omega^2 G^0({\bf r} - {\bf r}_0)G^0({\bf r}_0 - {\bf
r}')D(\omega)^{-1}.
\end{equation}
Here $G^0({\bf  r} - {\bf r}')$ is the Green's function of the polarization waves in
a system without defects.  Zeroes of $D(\omega)$,
\begin{equation}
\label{Localphonon}
D(\omega) = 1 - \gamma\omega^2\sum_k\frac{1}{\omega^2 - \omega_p^2({\bf k})},
\end{equation}
determine a frequency of local excitations in the absence of the interaction with the
electromagnetic
field.  In Eq. (\ref{Localphonon}) summation  over the first Brillouin zone  is assumed,  the
function $\omega_p({\bf k})$ is an initial dispersion law for the polarization waves. In the long
wavelength
approximation this function is taken in the form
\begin{equation}
\label{Phonondisp}
\omega^2_p (k)=\omega_0^2 - \Delta^2(ka)^2 ,
\end{equation}
which assumes a negative dispersion.  The parameter $\Delta$
in Eq. (\ref{Phonondisp}) gives the bandwidth of the spectrum of polarization waves.

Using the Green's function (\ref{Green}), one can obtain
the equation for the spatial Fourier transform of electromagnetic field, $E(k)$, in  the form 
\begin{equation}
\label{electricfield}
\omega^2\sum_{{\bf k}_1}\epsilon_{{\bf kk}_1}(\omega)E_{{\bf k}_1} = c^2k^2E_{{\bf
k}},
\end{equation}
where the dielectric constant, $\epsilon_{{\bf kk}_1}$,  is determined by the expression
\begin{equation}
\label{epsilon}
\epsilon_{{\bf kk}_1} = \displaystyle{\left(1 - \frac{d^2}{\omega^2 -
\omega_p^2(k)}\right)\delta_{{\bf kk}_1}} -
\displaystyle{\gamma\frac{d^2\omega^2\exp{[i({\bf k} - {\bf k}_1){\bf r}_0]}}{[\omega^2 -
\omega_p^2(k)][\omega^2 - \omega_p^2(k_1)]}}\frac{1}{D(\omega)}.
\end{equation}
The diagonal part of the dielectric constant corresponds to the interaction of electromagnetic waves with the excitations of the ideal structure and in the absence of the defects it would lead to the usual polaritons. The second, non-diagonal, term in Eq. (\ref{epsilon}) reflects the effect of the impurity. Its additional pole due to localized excitations of the crystal will be shown to bring on local polaritons.
Eq. (\ref{electricfield}) can be transformed into the following form
\begin{equation}
\label{electricfield1}
\left[(\omega^2 - \omega_l^2)(\omega^2 -\omega_p^2) - d^2\omega^2\right]E_k = \gamma 
d^2\omega^4\exp{(-i{\bf kr_0})}\displaystyle{\eta(E)}D(\omega),
\end{equation}
where 
$$
\eta(E)=\sum_{\bf k}\frac{E_k\exp{(-i{\bf kr_0}})}{\omega^2 - \omega_p^2(k)}
$$
and $\omega_l(k) = ck$ is the dispersion law of electromagnetic waves in the crystal in 
absence of the interaction with polarization waves. This is a uniform integral equation with the parameter $\eta(E)$ depending upon
all Fourier components, $E_k$, of the electromagnetic field. In order to find the electromagnetic field from this equation one has to transform it into the form
\begin{equation}
\label{elfreq}
\left(\gamma\omega^2 \sum_{\bf k}^{} \frac{\omega^2 - \omega^2_l (k)} 
{[\omega^2 - \omega^2_p (k)][\omega^2 - \omega^2_l (k)] -
d^2\omega^2} -1\right)\eta(E)=0.
\end{equation}
Parameter $\eta(E)$ has a non-zero value only if the frequency satisfies the equation
\begin{equation}
\label{Localfreq}
\gamma\omega^2 \sum_{\bf k}^{} \frac {\omega^2 - \omega^2_l (k)}
{[\omega^2 - \omega^2_p (k)][\omega^2 - \omega^2_l (k)] - d^2\omega^2}=1.
\end{equation}
Solutions to  Eq. (\ref{Localfreq}) determine new eigenfrequencies of the system which arise due to the
impurity. We show that corresponding eigenmodes are localized in the vicinity of
the impurity. The  parameter $\eta(E)$, though it is a function of $E$, is a free
parameter determined by initial or boundary conditions for these frequencies. It has a simple relationship with the polarization on the impurity site:
\begin{equation}
\label{eta}
 P({\bf r_0})= 4\pi d^2\eta(E) D^{-1}(\omega).
\end{equation}
If the frequency does not satisfy Eq. (\ref{Localfreq}), the parameter $\eta(E)$ is equal to zero and Eq. (\ref{electricfield1}) describes ordinary polaritons, which are not affected by the defect. As it follows from Eq.(\ref{eta}) the polarization of the defect in this case is also equal to zero, which means that the defect remains unexcited at all.
 
Eq. (\ref{Localfreq}) gives rise to real eigenmodes if the expression in the sum in this
equation does not have poles for real $\omega$ and $k$. It occurs only if $\omega$ belongs to
the gap between different polariton branches. This gap can exist if polarization waves have a
negative dispersion at small $k$, which we assumed in Eq. (\ref{Phonondisp}). Then
Eq. (\ref{elfreq}) describes two branches of polaritons with a gap
$\delta_p =\omega_2 - \omega_1$ between them. Here $\omega^2_2=\omega_0^2 + d^2$, and
$\omega_1^2\approx\omega^2_0 - \Delta \omega_0 d(a/c)$ is determined by the maximum
of the lower polariton branch. This maximum is an important general feature of the polariton
dispersion law in the presence of a bandgap. It has a topological origin and it always exists if
polarization excitations have a negative dispersion in the long wavelength limit. We show that this is responsible for the significant difference between properties of states within the
bandgap in our system and those within bandgaps caused by periodicity. 
 
For frequencies far from the polariton region, Eq. (\ref{Localfreq})  reduces to
the well known form of Lifshitz's equation \cite{phonon1,exciton1}, $D^{-1}(\omega)=0$, for the frequency of local phonon states. It is
interesting to note, however, that for frequencies outside of the polariton bandgap, Eq. 
(\ref{Localfreq}) gains an imaginary part due to poles for real $\omega$, which describe a
radiative decay of  local vibrations or excitons.

It is readily seen that for the frequencies inside the polariton gap the solution to the
Eq. (\ref{electricfield1}) has the form $E \propto E_0(\omega)\exp{[-\mid{\bf r} - {\bf
r}_0\mid]/l(\omega)}$,
where the localization length, $l$, can be derived from the polariton dispersion law. 
Near the lower boundary of the gap, its dependence upon frequency has the form 
$$
l^{-1}\propto \sqrt{\omega^2-\omega^2_1}/a\Delta,
$$
and near the upper boundary,
$$
l^{-1} \propto \frac{\omega_2}{c\sqrt{\omega_2^2-\omega_1^2}}
\sqrt{\omega_2^2-\omega^2}.
$$
The frequency dependence of the localization length according to these expressions is 
asymmetric. It falls off rapidly when the frequency moves away from the lower boundary and
increases slowly when the frequency approaches the upper boundary.
The amplitude of the field $E_0$ is given by the expression
\begin{equation}
E_0(\omega)\approx \gamma P({\bf r}_0)\frac{\omega^4a^2}{\Delta c^2\sqrt{\omega^2 -
\omega_1^2}},
\end{equation}
where we have used Eq. (\ref{eta}) in order to replace parameter $\eta(E)$ by $P({\bf r}_0)$. 

The frequency $\omega$ and the parameter $\gamma$ in this expression are connected to each other by dispersion eq. (\ref{Localfreq}). This equation containes the sum over all wave vectors including those from the edge of the Brilluen band and from both polariton branches. At small $k$ these branches are reduced into practically pure phonon and photon states. Photons with wavelengths comparable to a lattice parameter can not be treated in the framework of macroscopic theory used in this paper. However, it can be shown that contribution from photon-like branch in the region of large $k$ is negligible (one can see this merely by neglecting the coupling parameter $d$ in eq. (\ref{Localfreq})), and the only states contributing into the sum in eq. (\ref{Localfreq}) at large $k$ are pure phonons. This elimination of photon contribution at the edge of Brilluen band is due to the mechanical nature of the defects under consideration. If one considers a defect atom with the polarizability different from that of host crystals the numerators in eq. (\ref{Localfreq}) would have a form of $\omega^2 -\omega_p^2$.  Consequently, the major contribution to the sum at large $k$ would arise from pure photon states. The  sum with pure photon propagators, which arizes in this situation, is still assumed to be restricted by the first  Brilluien band, since higher wave numbers can be thought as included into dynamical matrix $L({\bf r}-{\bf r_1})$.   
Local phonon and exciton states in one- and two-dimensional systems exist for an arbitrarily small
value of $\gamma$. In 3-D systems, such states may appear only if  $\gamma >
\gamma_{cr}$. This difference is due to the behavior of the corresponding
integral for $\omega$ near the lower boundary of the gap: in 1-D and 2-D the integral
diverges whereas in 3-D it has a finite value, which determines the threshold for $\gamma$.
In the situation considered in this paper, the sum in Eq. (\ref{Localfreq})  diverges near the lower
boundary of the gap $\omega_1$ {\it even in a 3-D situation}. This singularity results from the
maximum of the lowest polariton branch at $\omega = \omega_1$.  In order to show how this
singularity occurs let us
rewrite the sum in Eq. (\ref{Localfreq}) as follows
\begin{equation}
\label{int}
\sum_{\bf k}^{} \frac{\omega^2 - \omega^2_l (k)} 
{\displaystyle{\left[\omega_2 - \Omega_1^2(k)\right]\left[\omega_2 - \Omega_2^2(k)\right]}},
\end{equation}
where $\Omega_1(k)$ and $\Omega_2(k)$ are dispersion laws for lower and upper polariton
branches respectively.
Near the lower edge of the gap,  $\Omega_1(k)^2$  can be presented as $\omega_1^2 - b(k^2 -
k_m^2)^2$. Since the corresponding integral has a pole of second order at $k=k_m$, it diverges when $\omega$
approches $\omega_1$. 

The  sum in Eq. (\ref{Localfreq}) cannot be evaluated in the long wavelength approximation since
details
of the spectrum of polarization waves for wave vectors near the boundary of the Brillouin zone
contribute
considerably to it.  In the vicinity of the frequency $\omega_1$, however, the only singular
part  of this sum becomes important.  Since the singularity occurs at the wave number $k_m
\approx (a\omega_0d)/(c\Delta)$, which is greater than the resonance wave number
($\omega_0/c$), but still far from the boundaries of the Brillouin zone, the singular part can
be evaluated in the long wavelength approximation. 

Neglecting the final contributions to the integral, one can obtain the following relation
between the frequency and the parameter $\gamma$
\begin{equation}
\label{gammazero}
\sqrt{\omega^2-\omega^2_1}\sim \gamma\frac{d\omega_0^3 a}{c\Delta^2}.
\end{equation}
This equation shows that 
in the harmonic approximation, {\it local polariton states arise for any value of $\gamma$ in 3-D systems}. This
result demonstrates the difference
between the two kind of gaps: those created by periodicity and by resonance interaction. This difference is due to that the density of states of excitations near the boundary of the first kind of gaps turns to zero, while in our case it diverges when the frequency approaches the lower boundary of the gap. 

According to Eq. (\ref{gammazero}) when $\gamma$ increases the frequency of the local state
tends away from the boundary, inside the gap. This guarantees the existence of
solutions at finite distance from the boundary at any finite value of $\gamma$ up to some
$\gamma_{cr}$ when the frequency becomes equal to the upper boundary of the gap
$\omega_2$. For larger $\gamma$, the frequency runs out of the gap and real valued solutions
of Eq. (\ref{Localfreq}) dissapear. The upper critical value, $\gamma_{cr}$, 
can
be calculated from Eq. (\ref{Localfreq}) with $\omega = \omega_2$. The leading contribution to
the integral at this frequency originates from $k$ at the boundary of the Brillouin zone.  In this
situation the integral reduces to the corresponding
integral in purely phonon or exciton problem.  However, it is interesting to note that the shift of local frequencies from
$\omega_2$ into the gap caused by a small deviation of $\gamma$ from $\gamma_{cr}$ is
strongly affected by the 
interaction with the electromagnetic field.  The value of such a shift,
$\delta\omega^2=\omega_2^2-\omega^2$, can be described by the following expression
\begin{equation}
\label{deltawup}
\frac{\delta\gamma}{\gamma} \propto
\gamma_{cr}\frac{2\pi^2\omega_2^5a^3}{c^3d^3}\sqrt{\delta\omega^2}+
\frac{\pi^2\omega_2^2}{\Delta^3d}\delta\omega^2,
\end{equation}
where $\delta\gamma=\gamma_{cr}-\gamma$. 
The first term in this equation determines $\omega(\gamma)$ near the upper
boundary, which is similar to the result obtained for local mechanical vibrations
\cite{phonon3}: $\omega_2 - \omega \propto \delta\gamma^2$. However, the coefficient of this
term is extremely small so that, in the immediate vicinity of the upper boundary we have a
crossover to the dependence $\omega_2 - \omega \propto \delta\gamma$, determined by the
second term on the right hand side of Eq. (\ref{deltawup}).  Hence, polariton effects result in a faster, 
{\it linear} shift of the
local frequency from the boundary in comparison with quadratic ``phonon-like" dependence.

States arising in the immediate vicinity of a boundary of a gap are sensitive to different
factors omitted in our model, such as anharmonicity or interactions with other subsystems.
This is true for our model as well as for local excitations with regular behavior near the
boundary of the gap. In general these factors result in the relaxation of all modes involved. This has a two-fold effect on the local states. First, they become
quasistationary with a finite lifetime. This time, however, is usually several orders of magnitude greater then the oscillation period of local modes (e.g., for hydrogen-localized vibrational modes in GaAs this ratio is $\sim 10^3$ \cite{time}). Since the relaxation of local polaritons is mainly determined by the decay of phonon or exciton components, one can expect their lifetime to be approximately of the same order of magnitude. Second, due to 
relaxation processes, the boundary between the continuous spectrum and the gap acquires a
finite width, $\Gamma$. This fact does not affect states deep inside of the gap, yet it
influences states in
the vicinity of the band edge. Due to the width $\Gamma$, these states appear on the background of the continuous
spectrum and they cannot then be considered as truly local states. Therefore, only states which
are father from the 
boundary than its relaxation width, $\Gamma$ can be considered to be local. This increases the
threshold for regular local states in
periodic systems and sets up a finite threshold for local polaritons.  This relaxation-induced
threshold can be
estimated from Eq. (\ref{gammazero}) provided that the threshold is small enough  that 
calculations of the integral in the main singular approximation still valid:
\begin{equation}
\label{threshold}
\gamma_{cr} \sim \frac{c\Delta^2}{ad\omega_0^2}\sqrt{\frac{\Gamma}{\omega_0}}.
\end{equation}
Numerical estimates  show that for
 excitons with high frequencies and small damping it is possible to
have  for $\gamma_{cr}  \sim  10^{-2}$, where we have used parameters for exciton polaritons
in KI from Ref. \cite{experiment}. We conlcude,
therefore, that one could observe local exciton-polariton states with frequencies
within the polariton gap, induced by
impurities with trap depth much smaller than for ordinary local excitons. 

Even if genuinely local states do not arise at small values of $\gamma$, this
does not mean that the singularity in the Eq. (\ref{Localfreq}) has no effect. It brings
on  so called ``virtual" or quasilocal states, which are also well known for phonon or exciton
excitations. Even though these resonances are not real eigenstates  of the system, they can
considerably affect thermodynamical or optical properties of the material. In the 
situation considered, one can expect that these resonance states could influence the shape of the absorption edge in the
vicinity of the lower boundary of the polariton gap. 

In this Letter we have shown that regular polar crystals with
impurities 
can be used to  create local states of polaritons with frequencies inside the polariton gap.  
Such local states in the case of a low concentration of impurities can manifest
themselves experimentally in the frequency dependence of the transmission coefficient inside the
forbidden gap. A similar effect has been observed for a single defect in a thin layer of a
photonic crystal \cite{Yablonovitch}. One should expect a peak in the transmission when
the frequency coincides with the frequency of a local state. It occurs
because an exponential tail of the incident field increases in the vicinity of the
impurity if the frequency of the field coincides with the eigenfrequency of the local
polaritons. These localized states provide one with as yet an unexplored yet opportunity to trap
the transverse electromagnetic field and to produce narrow optical filters. 

We wish to thank A.Z. Genack, S.A. Gredeskul and A.A. Maradudin  for useful comments on the
manuscript. We also benefitted from discussions with J.L. Birman, L.A. Pastur  and V. Podolsky.
This 
work was supported by the NSF under grant No.
DMR-9311605, by a CUNY collaborative grant, and by a PSC-CUNY research award.


\begin{thebibliography}{99}
\bibitem{phonon1} I.M. Lifshitz, Nuovo Cim. (Suppl A1) {\bf 3}, 591 (1956).
\bibitem{phonon2} A.A. Maradudin, E.W. Montroll, G.H. Weiss, and I.P. Ipatova {\it Theory of
Lattice
Dynamics in the Harmonic Approximation}, 2nd edition (Academic Press, New York, 1971).
\bibitem{phonon3} I.M. Lifshitz and A.M. Kosevich, in: {\it Lattice Dynamics} (Benjamin,
New
York, 1969), p. 53.
\bibitem{phonon4}  A.A. Maradudin, in: {\it Lattice Dynamics} (Benjamin, New York,
1969),
p. 1
\bibitem{exciton1} G.F. Koster and J.C. Slater. Phys. Rev. {\bf 95}, 1167 (1954)
\bibitem{exciton2}E.I. Rashba, Optika i Spectroskopia {\bf 2}, 586 (1957); R.E. Merrifield,
J. Chem. Phys. {\bf 38}, 920 (1963).
\bibitem{exciton3} V.M. Agranovich, {\it Theory of Excitons} (Nauka, Moscow, 1968).
\bibitem{magnon1} Yu. A. Izyumov, Adv. Phys. {\bf 14}, 569 (1965).
\bibitem{magnon2} R.J. Elliott. Comments on Solid State Physics, {\bf 1}, 118
(1968);  R.J. Elliott, J.A. Krumhansl, and P.L. Leath. Rev. Mod.
Phys., {\bf 46}, 465 (1974).
\bibitem{localreview} {\it Photonic Band Gaps and Localization}, edited by C.M. Soukoulis
(Plenum, New York, 1993).
\bibitem{Yablonovitch} E. Yablonovitch, T.J. Gmitter, R.D. Meade, A.M. Rappe, K.D. Brommer, and J.D. Joannopoulos. Phys. Rev. Lett.
{\bf 67}, 3380 (1991).
\bibitem{Joannopoulos} R.D. Meade, K.D. Brommer, A.M. Rappe, and J.D.
Joannopoulos, Phys. Rev. B, {\bf 44}, 13772, (1991).
\bibitem{Maradudin} A.A. Maradudin and J. Oitmaa, Solid State Communications, {\bf 7},
1143 (1969); D.L. Mills and A.A. Maradudin, Phys. Rev. B, {\bf 1}, 903 (1970).
\bibitem{Mills} R. Maddox and D.L. Mills, Phys. Rev. B, {\bf 11}, 2229 (1975).
\bibitem{polarlocal}Ze Cheng and Shi-Wei Gu, Phys. Rev. B {\bf 41}, 3128 (1990).
\bibitem{time} L.P. Wang, L.Z. Zhang, W.X. Zhu, X.T. Lu, and G.G. Qin. phys. stat. sol. (b), {\bf 158}, 113 (1990)
\bibitem{experiment}D. Fr\"{o}lich, P. K\"{o}hler, W. Nieswand, and T. Rappen. Phys. Rev. B,
{\bf 43}, 12590 (1991)
\end{thebibliography}
\end{document}